\documentclass[12pt]{article}
\usepackage{graphicx}
\usepackage{calrsfs}
\usepackage{geometry}                		
\usepackage{graphicx}				
\usepackage{amssymb}
\usepackage[pdfstartview=FitH,
colorlinks=true,linkcolor=red,anchorcolor=black,citecolor=blue,urlcolor=black,filecolor=blue]{hyperref}

\usepackage{amsmath,amssymb,titling,authblk}
\usepackage{amscd}
\usepackage{caption}
\usepackage{etoolbox}
\usepackage{makeidx}
\include{epsf}
\usepackage[toc,page]{appendix}
\usepackage{amsfonts}
\usepackage{authblk} 
\usepackage{tcolorbox}
\usepackage{sectsty}
\usepackage{amsmath,amssymb,epsfig}
\usepackage{amscd}
\usepackage{amsthm}
\usepackage{braket}
\usepackage{amsmath}
\usepackage{xcolor}
\usepackage{amssymb}
\usepackage{mathrsfs}
\usepackage{setspace}
\usepackage{url}
\usepackage{dsfont}
\usepackage[applemac]{inputenc}
\usepackage[english]{babel}
\usepackage{enumitem} 
\usepackage{verbatim}
\usepackage{slashed}
\usepackage[]{latexsym}
\usepackage{mathtools}
\usepackage{subcaption}
\usepackage{ mathrsfs }
\usepackage{braket}
\usepackage{caption}
\usepackage{cite}
\usepackage{dsfont}
\usepackage{mathpazo}
\usepackage{hyperref}
\hypersetup{
pdftitle={},%
pdfauthor={},%
pdfsubject={},%
pdfkeywords={},%
colorlinks=true,%
linkcolor=blue,%
citecolor=red,%
linktocpage=true,%
pageanchor=true
}

\newcommand{\be}{\begin{equation}}
\newcommand{\ee}{\end{equation}}
\def\beqa{\begin{eqnarray}}
\def\eeqa{\end{eqnarray}}
\def\bean{\begin{eqnarray*}}
\def\eean{\end{eqnarray*}}

\newcommand{\dd}{{\mathrm{d}}}

\newcommand{\ii}{\mathrm{i}} 
\theoremstyle{definition}

\renewenvironment{thebibliography}[1]
         {\section*{References}\frenchspacing\small
          \begin{list}{[\arabic{enumi}]}
         {\usecounter{enumi}\parsep=2pt\topsep 0pt
         \settowidth{\labelwidth}{[#1]}
         \leftmargin=\labelwidth\advance\leftmargin\labelsep
         \rightmargin=0pt\itemsep=1pt\sloppy}}{\end{list}}

 \numberwithin{equation}{section}

\title{\textbf{ {Hamiltonian analysis in Lie-Poisson gauge theory}}\vspace{0.5cm}}
\date{}

\author[1]{Francesco Bascone}
\author[1,2]{Maxim Kurkov}
\affil[ ]{}
\affil[1]{\textit{\footnotesize INFN-Sezione di Napoli, Complesso Universitario di Monte S. Angelo Edificio 6, via Cintia, 80126 Napoli, Italy.}}
\affil[2]{\textit{\footnotesize Dipartimento di Fisica ``E. Pancini'', Universit\`a di Napoli Federico II, Complesso Universitario di Monte S. Angelo Edificio 6, via Cintia, 80126 Napoli, Italy.}}
\affil[ ]{}
\affil[ ]{\footnotesize e-mail: \texttt{francesco.bascone@na.infn.it, max.kurkov@gmail.com}}
\begin{document}
\maketitle
\begin{abstract}

{Lie-Poisson gauge formalism provides a semiclassical description of noncommutative $U(1)$ gauge theory with  Lie algebra type noncommutativity. 
 Using the Dirac approach to constrained Hamiltonian systems, we focus on a class of Lie-Poisson gauge models, which exhibit an admissible Lagrangian description.
 The underlying noncommutativity is supposed to be purely spatial. Analysing the constraints, we demonstrate 
 that these models have as many physical degrees of freedom as there are present in  the  Maxwell theory. 
 }

\end{abstract}

\newpage
\tableofcontents

\section{Introduction}\label{secintroduction}
{
Noncommutative geometric structures arise in various areas of theoretical and mathematical physics~\cite{Castellani:2000xt,Lizzi:2018dah,Devastato:2019grb}. Considerable attention has been paid to
noncommutative field theories~\cite{SzaboReview}, in particular, to noncommutative gauge models~\cite{WalletReview,Meier:2023lku}. The present paper is devoted to the Poisson gauge theory (PGT) designed in~\cite{Kupriyanov:2021aet}. The PGT formalism provides a semiclassical approximation of the novel approach to noncommutative gauge theory, proposed in~\cite{Kupriyanov:2020sgx}. A profound connection of the PGT with  $L_{\infty}$ structures was discussed in~\cite{Blumenhagen:2018kwq,Kupriyanov:2021cws,Abla:2022wfz}. Of course, there other approaches to semiclassical limits of noncommutative gauge models, e.g. the one of~\cite{Jurco:2000fs}, see~\cite{Kupriyanov:2023gjj} for detailed comparison.

 Various noncommutativities, including the canonical, the 
$\lambda$-Minkowski and the $\kappa$-Minkowski cases were studied in the context of the PGT~\cite{Abla:2023odq,Kupriyanov:2020axe,Kurkov:2021kxa}.  We shall limit ourselves to noncommutativities of Lie algebra type, which yield the so called Lie-Poisson gauge models~\cite{Kupriyanov:2023gjj}.

For a certain class of noncommutativities the Lie-Poisson gauge formalism exhibits an admissible Lagrangian formulation~\cite{Kupriyanov:2023gjj}. The corresponding dynamics can be obtained from a local gauge-invariant classical action, recovering the standard Maxwell action at the commutative limit. However, the corresponding Hamiltonian analysis has never been performed, and the present research is addressed to fill this gap.  Such a study sheds light onto the number of \emph{physical degrees of freedom} (d.o.f.). 
Indeed, the seminal treatment of P.A.M. Dirac in~\cite{Dirac} shows that any
gauge invariance necessarily yields constraints in the phase space. 
Therefore the number $N$ of physical d.o.f. is always smaller than the number $n_1$ of generalised coordinates\footnote{In field theoretical models the values of  fields at different spatial points are considered as the generalised coordinates, thus there are infinitely many d.o.f. In this case one has to count  the numbers $N$, $n_1$, $n_2$ and $n_3$ per each spatial point, which are finite. } present in a gauge model~\cite{Henneaux:1990au},
\be
N = n_1 - n_2 - \frac{1}{2}n_3 , \label{numberdeof}
\ee  
where $n_2$  and $n_3$ stand for the total numbers of the first and the second class constraints respectively. For example, in the $n$-dimensional $U(1)$ gauge theory, $n_1=n$, $n_2=2$ and $n_3=0$, so we end up with $n-2$ physical d.o.f.

It is known that the space-time noncommutativity  may have drastic effects on gauge models,  e.g. creating the Gribov ambiguity~\cite{Canfora:2015nsa,Kurkov:2017tyn}. 
One may wonder whether noncommutative deformations affect the number of physical d.o.f. 
 In the present paper we answer this question for the Lie-Poisson gauge theory, which exhibits an admissible Lagrangian formulation in the sense of~\cite{Kupriyanov:2023gjj}. Limiting ourselves  
to purely spatial noncommutativities, we shall analyse the constraints within the Dirac formalism. We shall demonstrate that \emph{the noncommutative deformations do not change the number of physical d.o.f.}

This paper is organised as follows. In Sec.~\ref{SecRev} we set the notations and briefly review  the Poisson gauge formalism and its relation to the noncommutative geometry. The main part of the paper is Sec.~\ref{CostrHam}, where we elaborate the constraints.  Sec.~\ref{Conclu} contains our conclusions. Various technicalities are summarised in Appendices~\ref{appA} and~\ref{appB}.

\section{Basics of the (Lie-) Poisson gauge formalism \label{SecRev}}
First we discuss the noncommutative geometric origin of the (Lie-) Poisson gauge theory. After that we introduce the main technical elements of the formalism, which will be used in the subsequent analysis of Hamiltonian constraints in Sec.~\ref{CostrHam}.
\subsection*{a. Poisson gauge theory and noncommutative geometry}
Let $\mathcal{M}$ be a $n$-dimensional manifold, representing the space-time. The local coordinates on $\mathcal{M}$ are denoted through $x^{\mu}$, $\mu=0,...,n-1$. The noncommutative structure of $\mathcal{M}$ is described by the Kontsevich star product,
\begin{equation}
f(x) \star g(x)=f(x) \cdot g(x)+\frac{\ii}{2}\{f(x),g(x) \}+ \dots, \quad f,g \in C^{\infty}(\mathcal{M}),
\end{equation}
where the Poisson bracket 
\begin{equation} \label{mpb}
\{f,g \}
=\Theta^{\mu \nu}\partial_{\mu}f \, \partial_{\nu}g, \quad f,g \in C^{\infty}(\mathcal{M}),
\end{equation}
is related to the Poisson bivector $\Theta^{\mu\nu}$, and  dots stand for  the remaining terms, which  contain higher derivatives of $f$ and $g$. 

Let $A_{\mu}(x)$ be a $U(1)$ gauge connection on $\mathcal{M}$.
According to the proposal of~\cite{Kupriyanov:2020sgx},  noncommutative deformations of the $U(1)$ gauge theory must satisfy the two postulates listed below. 
\begin{itemize}
\item{Infinitesimal gauge transformations must close the non-commutative gauge algebra, 
\begin{equation}\label{eqgaugeclosure}
\left[\delta_f, \delta_g\right] A_{\mu}(x)=\delta_{-\mathrm{i}[f(x), g(x)]_{\star}} A_{\mu}(x), 
\end{equation}
where 
\begin{equation} \label{starcommut}
[f(x), g(x)]_{\star}:=f(x) \star g(x)-g(x) \star f(x), \qquad f,g\in\mathcal{C}^\infty(\mathcal{M}). 
\end{equation}
}
\item{At the commutative limit  the infinitesimal gauge transformations must recover the standard Abelian gauge transformations,
\begin{equation}\label{eqcommlimit}
\lim _{\Theta \rightarrow 0} \delta_f A_{\mu}(x)=\partial_{\mu} f(x).
\end{equation}}
\end{itemize}
In the semiclassical approximation,
\begin{equation}
[f(x), g(x)]_{\star}  \simeq  \ii \{f(x),g(x) \},
\end{equation}
 so the full non-commutative algebra~\eqref{eqgaugeclosure} reduces to the \emph{Poisson gauge algebra},
\begin{equation}\label{eqpoissongaugeclosure}
\left[\delta_f, \delta_g\right] A_{\mu}=\delta_{\{f, g\}} A_{\mu}, \qquad f,g\in \mathcal{C}^\infty(\mathcal{M}).
\end{equation}
By definition, the  \emph{Poisson gauge theory}  is a field theoretical model with the gauge transformations, closing the algebra~\eqref{eqpoissongaugeclosure} and obeying the commutative limit~\eqref{eqcommlimit}. In other words, the PGT is a semiclassical approximation of the noncommutative $U(1)$ gauge theory, constructed along the lines of~\cite{Kupriyanov:2020sgx}.

In this paper we shall focus on a special class of extremely useful Poisson structures, which are linear in the space-time coordinates,
\begin{equation}
\Theta^{\mu \nu}(x)=f_{\alpha}^{\mu \nu}\, x^{\alpha}.
\end{equation}
The tensor $\Theta$ is a Poisson bivector iff
\begin{equation}
f_{\mu}^{\tau \alpha} f_{\alpha}^{\beta \delta}+f_{\mu}^{\beta \alpha} f_{\alpha}^{\delta \tau}+f_{\mu}^{\delta \alpha} f_{\alpha}^{\tau \beta}=0,
\end{equation}
what is nothing but the Jacobi identity for the structure constants of a Lie algebra.
In~\cite{Kupriyanov:2023gjj} the PGTs of this type were called \textit{Lie-Poisson gauge theories}. From now on we limit ourselves to the Lie-Poisson gauge models only.

\subsection*{b. General construction}
The key technical elements of the formalism are the field-dependent matrices  $\gamma(A)$ and $\rho(A)$, which solve the master equations, 
\be
\left\{\begin{array}{l}
\gamma_{ \mu }^{ \nu } \partial^{ \mu }_A \gamma^{ \xi}_{ {\lambda}} - \gamma^{ \xi}_{ \mu } \partial_A^{ \mu } \gamma^{ \nu }_{\lambda} 
  = \gamma^{ \mu }_{ \lambda} f_{ \mu }^{ \nu  \xi} \\
\gamma^\nu_{\lambda} \partial_A^{\lambda}\rho_\xi^\mu \,+ \rho_\xi^{\lambda} \partial_A^\mu\gamma_{\lambda}^\nu   = 0  
\end{array}\right.,  \qquad \partial_A^{\lambda} \equiv \frac{\partial}{\partial A_{\lambda} },  \label{master}
\ee
and tend to the identity matrix at the commutative limit, 
\be
\lim_{\Theta\to 0} \gamma = \mathbb{I},\qquad \lim_{\Theta\to 0} \rho = \mathbb{I}.  \label{mcomlim}
\ee
The explicit solutions of these equations  were obtained in~\cite{Kupriyanov:2021cws} and~\cite{Kupriyanov:2022ohu} for arbitrary structure constants $f_{ \mu }^{ \nu  \xi}$.

The infinitesimal Poisson gauge transformations, which close the gauge algebra~\eqref{eqpoissongaugeclosure} and respect the commutative limit~\eqref{eqcommlimit}, can be constructed using the matrix $\gamma$~\cite{Kupriyanov:2019cug, Kupriyanov:2020sgx},
\begin{equation}\label{eqgaugetransfgeneral}
\delta_f A_{\mu}(x)=\gamma_{\mu}^{\nu}(A(x)) \partial_{\nu} f(x)+\left\{A_{\mu}(x), f(x)\right\}, \qquad f\in  \mathcal{C}^\infty(\mathcal{M}).
\end{equation}
Another important constituent of the PGT  formalism is the deformed field strength $\mathcal{F}_{\mu\nu}$, which transforms in a gauge-covariant manner upon these transformations,
\be
\delta_f \mathcal{F}_{\mu \nu}(x)=\left\{\mathcal{F}_{\mu \nu}(x), f(x)\right\},\qquad f\in\mathcal{C}^\infty(\mathcal{M}),
\ee
and reproduces the  standard Abelian field strength at the commutative limit,
\be
\lim _{\Theta \rightarrow 0} \mathcal{F}_{\mu \nu}(x)=\partial_{\mu} A_{\nu}(x)-\partial_{\nu} A_{\mu}(x) =: F_{\mu\nu} .
\ee
The expression for this field strength was obtained in~\cite{Kupriyanov:2021aet} in terms of $\gamma$ and $\rho$ in the following form,
\begin{equation}\label{eqfieldstrength}
\mathcal{F}_{\mu \nu}(x)=\rho_{\mu}^{\alpha}(A(x)) \rho_{\nu}^{\beta}(A(x))\left(\gamma_{\alpha}^{\delta}(A(x)) \partial_{\delta} A_{\beta}-\gamma_{\beta}^{\delta}(A(x)) \partial_{\delta} A_{\alpha}+\left\{A_{\alpha}, A_{\beta}\right\}\right).
\end{equation}

From now on we assume that $\mathcal{M} = \mathbb{R}^{n}$, and the coordinates $x^{\mu}$ are Cartesian.  We also assume that $\mathcal{M}$ is equipped with the  Minkowski metric,  
\be
\eta_{\mu\nu} = \mathrm{diag}\big(+1,\,-1,..., -1\big)_{\mu\nu} = \eta^{\mu\nu},
\ee
which will be used to rise and to lower indices. The spatial coordinates $x^j$, 
$j=1,...,n-~1$, will be labeled by Latin letters. 

Iff the structure constants $f_{\mu}^{\nu\xi}$, defining the non-commutativity, obey the compatibility relation
\be 
f_{\mu}^{\mu\xi} = 0, \label{compatcomb}
\ee
an admissible gauge-invariant classical action for $A$ can be constructed in a simple  manner, 
\begin{equation}
S[A]=\int_{\mathcal{M}} \dd^n x  \,\mathscr{L}, \label{action}
\end{equation}
where the local gauge-covariant Lagrangian density\footnote{$\delta_f \mathscr{L}=\{\mathscr{L}, f \}$} is given by
\begin{equation}
\mathscr{L}=-\frac{1}{4} \mathcal{F}_{\mu \nu} \mathcal{F}^{\mu \nu}, \label{Lade}
\end{equation}
see~\cite{Kupriyanov:2023gjj} for details.
At the commutative limit the expression~\eqref{action} recovers the standard Maxwell action.
\subsection*{c. Purely spatial noncommutativity}
Consider a Lie algebra type noncommutativity, which does not involve the time variable $x^0$,
\be
f^{0\mu}_\nu = 0 = f^{\mu\nu}_0, \qquad \forall \mu,\nu.  
\ee
One can easily see that in this case the solutions of the master equations~\eqref{master} can be chosen in the following form,
\begin{equation}\label{rhogammanoncommspatial}
\rho_{\mu}^0=\delta_{\mu}^0=  \gamma_{\mu}^0, \qquad\qquad \rho^{\mu}_0=\delta^{\mu}_0=  \gamma^{\mu}_0, \qquad \forall \mu,
\end{equation}
whilst the $n-1$ by $n-1$ matrices $\gamma^i_j$ and $\rho^i_j$ do not depend on $A_0$ and obey the $n-1$-dimensional version of the relations~\eqref{master} and~\eqref{mcomlim}.

In the next section we perform a Hamiltonian constraint analysis of the Poisson gauge theory assuming the noncommutativity to be purely spatial, and considering the matrices $\gamma$ and $\rho$ of the structure, described above. 
Apart from that, we shall assume that the compatibility condition~\eqref{compatcomb} is satisfied, i.e. an admissible Lagrangian formulation is available.

\section{Hamiltonian constraint analysis \label{CostrHam}}
Below we derive and analyse the primary and the secondary constraints. We shall see that these constraints are of the first class, whilst no other constraints are present. 
\subsection*{a. Canonical Hamiltonian and primary constraints}
Consider a Lagrangian functional,
\be
L[A,\dot{A}]= \int_{\Sigma(x^0)}\dd^{n-1} x \, \mathcal{L},
\ee
where the Lagrangian density is defined by Eq.~\eqref{Lade}, and $\Sigma(x^0) =\mathbb{R}^{n-1}$ stands for a space-like hyperplane of constant time $x^0$.
Hereafter we use a dot notation for the time derivative, in particular,
$
\dot{A}_{\mu}  = \frac{\partial A_{\mu}}{\partial x^0}.
$
Introducing the momenta, conjugated to $A_{\mu}(x)$,
\begin{equation}\label{momentacalc}
\Pi^{\mu}:=\frac{\delta L}{\delta \dot{A}_{\mu}}= \frac{\partial \mathcal{L}}{\partial \dot{A}_{\mu}} =-\frac{1}{2}\mathcal{F}^{\alpha \beta}\frac{\partial\mathcal{F}_{\alpha \beta}}{\partial \dot{A}_{\mu}}= -\rho_{\alpha}^{\mu}\mathcal{F}^{0 \alpha},
\end{equation}
we immediately get the primary constraints,
\be \label{firsconstr}
\phi^1(x) \thickapprox 0,  \qquad \phi^1(x) := \Pi^0(x),
\ee
which are identical to the corresponding commutative expressions. Following Dirac we use the symbol $\thickapprox$ to indicate an equality  on the constraint surface. 

One can easily check that no other primary constraints are present, and the generalised velocities $\dot{A}_{j}$ can be expressed through the phase space coordinates as follows,  
\begin{equation} \label{velo}
\dot{A}_k(x)= \big[\rho^{-1}\big(A(x)\big)\big]^j_k\,  \big[\rho^{-1}\big(A(x)\big)\big]_i^j\, \Pi^i(x) +  Q^\ell_k(x)\,\partial_{\ell}A_0(x),
\end{equation}
where we introduced the notation
\be
Q^\ell_k(x) := \big[\gamma\big(A(x)\big)\big]^\ell_k -\Theta^{\ell p}(x)\partial_p A_k(x). \label{Qdef}
\ee

Defining the canonical Hamiltonian in a standard way, 
\begin{equation}\label{eqhamiltoniangen}
H_{\mathrm{can}}=\int_{\Sigma(x^0)} \dd^{n-1}x \left[\Pi^{\mu}\partial_0 A_{\mu}-\mathscr{L} \right],
\end{equation}
and taking into account the relations~\eqref{firsconstr} and~\eqref{velo}, through a sequence of simple calculations, which involve integrations by parts\footnote{Of course, we assume that all the fields decay sufficiently fast at the spatial infinity. }, we arrive at
\begin{equation}
H_{\mathrm{can}}= \int_{\Sigma(x^0)} \dd^{n-1}x\, \mathcal{H}_{\mathrm{can}},
\end{equation}
where the canonical Hamiltonian density is given by
\be
\mathcal{H}_{\mathrm{can}} =\frac{1}{4}\mathcal{F}^{ij} \mathcal{F}_{ij}
+\frac{1}{2}(\rho^{-1} )^i_j (\rho^{-1} )_k^i \Pi^j \Pi^k
-A_0 \,\partial_{\ell} \big[Q^\ell_k\,\Pi^k\big]. 
\ee 
At the commutative limit one recovers the  standard textbook expression~\cite{Ashok},
\be
\lim_{\theta\to 0}H_{\mathrm{can}}=\int_{\Sigma(x^0)} \dd^{n-1}x \left(\frac{1}{4}F^{ij} F_{ij}+\frac{1}{2}\Pi^k \Pi^k-A_0\, \partial_{k}\Pi^k \right).
\ee

Hereafter we use the following notation for the equal-time Poisson bracket of two arbitrary functionals $F$ and $G$ of the fields $A_{\mu}$ and $\Pi^{\mu}$ on $\Sigma(x^0)$,
\begin{equation} \label{thPB}
\{F, G\}_{\mathrm{p.s.}}:=\int_{\Sigma(x^0)} \dd^{n-1} z\left(\frac{\delta F }{\delta A_\mu(z)} \frac{\delta G}{\delta \Pi^\mu(z)}-\frac{\delta F}{\delta \Pi^\mu(z)} \frac{\delta G}{\delta A_\mu(z)}\right).
\end{equation}
We introduced the subscript ``p.s.", meaning the ``phase space", in order to distinguish~\eqref{thPB} from the Poisson bracket~\eqref{mpb} on $\mathcal{M}$, which defines the noncommutative gauge algebra~\eqref{eqpoissongaugeclosure}. In particular,
\begin{eqnarray}
\{A_{\mu}(x), \Pi^{\nu}(y)\}_{\mathrm{p.s.}}  &=& \delta_{\mu}^{\nu} \delta^{n-1}(x-y),\nonumber\\
\{A_{\mu}(x), A_{\nu}(y)\}_{\mathrm{p.s.}}  &=&0 =\{\Pi^{\mu}(x), \Pi^{\nu}(y)\}_{\mathrm{p.s.}} , 
\end{eqnarray}
where $x^0 = y^{0}$, and $$\delta^{n-1}(x-y) = \delta(x^1-y^1)\cdot ... \cdot \delta(x^{n-1}-y^{n-1}),$$ is a $n-1$-dimensional delta-function. 

Obviously,
\be
\{\phi^1(x),\phi^1(y)\}_{\mathrm{p.s.}} = 0. \label{phi11com}
\ee

\subsection*{b. Secondary constraints}

Consider the stability condition of the primary constraints~\eqref{firsconstr},
\be
\dot{\phi}^1(x)\thickapprox \{\phi^1(x),  H_{\mathrm{can}} \}_{\mathrm{p.s.}}   \thickapprox 0.
\ee
The relevant Poisson bracket reads, 
\begin{equation}
\{\Pi^0(x), H_{\mathrm{can}} \}_{\mathrm{p.s.}}  =- \frac{\delta H_{\mathrm{can}}}{\delta A_0(x)} = \partial_{\ell}\left[Q_k^{\ell}(x)\,\Pi^k(x)\right] ,
\end{equation}
so we are led to the secondary constraints
\begin{equation} \label{seccostr}
\phi^2(x)\thickapprox 0,  \qquad\phi^2(x):= \partial_k \left[Q^k_i(x)\, \Pi^i (x)\right],
\end{equation}
which reduce to the Gauss law  $\partial_k \Pi^k\thickapprox0$ at the commutative limit. In Appendix~\ref{appB} we demonstrate that
\be
\{\phi^2(x),\phi^2(y)\}_{\mathrm{p.s.}} = -\{\phi^2(x),\delta^{n-1}(x-y)\}, \label{twotwocomm}
\ee
therefore these constraints Poisson-commute with each other on the constraint surface,
\be
\{\phi^2(x),\phi^2(y)\}_{\mathrm{p.s.}} \thickapprox 0 . \label{twotwocommBis}
\ee

Continuing the Dirac procedure, we write down the primary Hamiltonian,
\be
H_{\mathrm{p}}= \int_{\Sigma(x^0)} \dd^{n-1}x\, \big(\mathcal{H}_{\mathrm{can}} + \lambda_1 \, \Pi^0  \big),
\ee
where $\lambda_1(x)$ stands for the Lagrangian multiplier, associated with the primary constraint~\eqref{firsconstr}.
The stability condition of the secondary constraint, 
\be
\dot{\phi}^2(x)\thickapprox \{\phi^2(x),  H_{\mathrm{p}} \}_{\mathrm{p.s.}}  \thickapprox 0,
\ee
is more subtle. As we mentioned at the end of the previous section, by construction $\gamma^i_j$ does not depend on $A_0$, therefore
\be
\{{\phi}^1(x),{\phi}^2(y)\}_{\mathrm{p.s.}} = 0, \label{phi12com}
\ee
and hence
\be 
\{\phi^2(x),  H_{\mathrm{p}} \}_{\mathrm{p.s.}}  \thickapprox \{\phi^2(x),  \hat{H}_{\mathrm{can}} \}_{\mathrm{p.s.}} 
= (\mathrm{I}) + (\mathrm{II}). \label{oneandtwo}
\ee
In this formula
\begin{eqnarray} \label{secondcheck}
(\mathrm{I}) &:=& \partial_k \left[ Q^k_j(x)\{\Pi^j(x), \hat{H}_{\mathrm{can}} \}_{\mathrm{p.s.}} \right],\nonumber\\
(\mathrm{II}) &:=& \partial_k \left[ \Pi^j(x)\{Q^k_j(x),\hat{H}_{\mathrm{can}}\}_{\mathrm{p.s.}} \right],
\end{eqnarray}
and
\begin{eqnarray}
\hat{H}_{\mathrm{can}} &:=& {H}_{\mathrm{can}} + \int_{\Sigma(x^0)} \dd^{n-1}x \,A_0(x)\, \phi^2(x) 
\nonumber\\
&=& \int_{\Sigma(x^0)} \dd^{n-1}x \,\bigg(\frac{1}{4}\mathcal{F}^{ij} \mathcal{F}_{ij}
+\frac{1}{2}\,(\rho^{-1} )^i_j (\rho^{-1} )_k^i \Pi^j \Pi^k\bigg).
\end{eqnarray}
Below we analyse the contributions $(\mathrm{I})$ and $(\mathrm{II})$ separately.

First we focus on the expression~$(\mathrm{I})$. 
In order to simplify our calculations we perform the following trick. We notice that
\begin{equation} \label{unObserv}
\{\Pi^j(x), \hat{H}_{\mathrm{can}} \}_{\mathrm{p.s.}} 
=- \frac{\delta \hat{H}_{\mathrm{can}}}{\delta A_j(x)}   
 = -\partial_A^j (\rho^{-1})^i_p(\rho^{-1})_k^i \Pi^p(x) \Pi^k(x)+\frac{\delta S_{n-1}[A,x^0] }{\delta A_j(x)},
\end{equation}
where for any fixed value of the time-coordinate $x^0$ the expression
\be
S_{n-1}[A,x^0]  := -\frac{1}{4} \int_{\Sigma(x^0)} \dd^{n-1}x\,  \mathcal{F}_{ij} \mathcal{F}^{ij} \label{dm1action}
\ee
has the structure of the Euclidean action of a $n-1$ dimensional PGT.  As we explained in detail at the end of the previous section, our noncommutativity is essentially spatial. Therefore upon the 
$n-1$-dimensional gauge transformations
\begin{equation}\label{eqgaugetransfgeneralDM1} 
\delta_{\tilde f} A_{i} =\gamma_{i}^{j}(A) \partial_{j} \tilde f+\left\{A_{i}, \tilde{f} \right\}, \qquad \tilde{f}\in \mathcal{C}^{\infty}(\Sigma(x^0))
\end{equation}
closing the $n-1$-dimensional Poisson gauge algebra
\be \label{Poisdm1}
\left[\delta_{\tilde{f}}, \delta_{\tilde{g}}\right] A_{j}=\delta_{\{\tilde{f}, \tilde{g}\}} A_{j}, \qquad  \tilde{f},\tilde{g}\in \mathcal{C}^{\infty}(\Sigma(x^0))
\ee
the spatial components of the deformed field strength~\eqref{eqfieldstrength} transform in a gauge covariant way,
\be
\delta_{\tilde f} \mathcal{F}_{ij} =\left\{\mathcal{F}_{ij}, \tilde{f}\right\},  \qquad \tilde{f}\in \mathcal{C}^{\infty}(\Sigma(x^0)).
\ee
Thus the  $n-1$-dimensional action~\eqref{dm1action} remains invariant upon the transformations~\eqref{eqgaugetransfgeneralDM1},
and the associated Noether identity reads 
\be \label{Notiden}
\partial_k \left[ Q^k_j(x)\, \frac{\delta S_{n-1}[A,x^0] }{\delta A_j}\right] = 0, 
\ee
see  Appendix A for details. Substituting the relation~\eqref{unObserv} in Eq.~\eqref{secondcheck}, and by using the identity~\eqref{Notiden},  we get
\begin{eqnarray}
(\mathrm{I}) &=& -\partial_k\left[ Q^k_j(x)\,\partial_A^j (\rho^{-1})^i_p(\rho^{-1})_q^i \,\Pi^p \Pi^q \right]  \nonumber\\
 &=&- \frac{1}{2}
 \{(\rho^{-1})^i_p(\rho^{-1})_q^i, \Pi^p \Pi^q\}
 - \partial_k \left[\gamma^k_j \,\Pi^p \Pi^q\, (\rho^{-1})_q^i \,\partial_A^j (\rho^{-1})^i_p  \right]. \label{onefinal}
\end{eqnarray}

Now we elaborate the expression~$(\mathrm{II})$.
One can easily see that
\begin{eqnarray}
\{Q_j^k(x),\hat{H}_{\mathrm{can}}\}_{\mathrm{p.s.}} &=&  \int_{\Sigma(x^0)}\dd^{n-1} z \,\, \frac{\delta \hat{H}_{\mathrm{can}}}{\delta \Pi^r(z)} \frac{\delta  Q_j^k(x)
 }{\delta A_r(z)} \nonumber\\
&=& \Pi^b(x)  (\rho^{-1})^i_r  (\rho^{-1})^i_b \partial_A^r\gamma^k_j + \Theta^{pk}(x)\, \partial_p \big[(\rho^{-1})^i_j
(\rho^{-1})^i_b \Pi^b  
\big], \nonumber
\end{eqnarray}
therefore
\be
(\mathrm{II}) =\partial_k\Big[\Pi^j \Pi^b  (\rho^{-1})^i_r  (\rho^{-1})^i_b \partial_A^r\gamma^k_j \Big]
-\{\Pi^j,(\rho^{-1})^i_j
(\rho^{-1})^i_b \Pi^b  \},
\ee
where we took into account the compatibility relation $\partial_k  \Theta^{pk} = 0$.

It is also easy to check that
\begin{eqnarray}
\{\Pi^j,(\rho^{-1})^i_j
(\rho^{-1})^i_b \Pi^b  \}  &=& \underbrace{\{\Pi^j,\Pi^b  \} (\rho^{-1})^i_j
(\rho^{-1})^i_b }_0 +  \Pi^b\{\Pi^j,\underbrace{(\rho^{-1})^i_j
(\rho^{-1})^i_b}_{\mbox{sym. in $j$ and $b$}}  \} \nonumber\\
&=& \frac{1}{2} \big(\Pi^b\{\Pi^j,(\rho^{-1})^i_j
(\rho^{-1})^i_b + \Pi^j\{\Pi^b,(\rho^{-1})^i_j
(\rho^{-1})^i_b \}\big) \nonumber\\
&=& \frac{1}{2} \{\Pi^j\Pi^b,(\rho^{-1})^i_j
(\rho^{-1})^i_b  \},
\end{eqnarray}
thus
\be \label{twofinal}
(\mathrm{II}) =\partial_k\Big[\Pi^p \Pi^q  (\rho^{-1})^i_j  (\rho^{-1})^i_q \partial_A^j\gamma^k_p \Big] - \frac{1}{2} \{\Pi^p\Pi^q,(\rho^{-1})^i_p
(\rho^{-1})^i_q  \},
\ee
where we renamed the mute indices in a suitable way.

Substituting the expressions~\eqref{onefinal} and~\eqref{twofinal} in Eq.~\eqref{oneandtwo} we immediately obtain
\be
\{\phi^2(x),  H_{\mathrm{p}} \}_{\mathrm{p.s.}} = \partial_k\Big[\Pi^p \Pi^q  (\rho^{-1})^i_q \Big ((\rho^{-1})^i_j  \partial_A^j\gamma^k_p 
-  \gamma^k_j  \partial_A^j (\rho^{-1})^i_p   \Big)\Big].
\ee
According to~\cite{Kupriyanov:2022ohu}, the (three-dimensional) master equation for $\rho$ can be rewritten in the following way,
\be
(\rho^{-1})^i_j  \partial_A^j\gamma^k_p 
-  \gamma^k_j  \partial_A^j (\rho^{-1})^i_p    = 0,
\ee
hence
\be
\{\phi^2(x),  H_{\mathrm{p}} \}_{\mathrm{p.s.}} = 0,
\ee
i.e. there are no other constraints. The relations~\eqref{phi11com}, \eqref{twotwocommBis} and~\eqref{phi12com} imply that the constraints  $\phi^1(x)$ and $\phi^2(x)$ are of the first class. Since the second class constraints are absent, by setting $n_1 = n$, $n_2=2$ and $n_3=0$ in Eq.~\eqref{numberdeof},
we end up with  
\be
N=n-2 
\ee
physical degrees of freedom. 
 }
 
 \subsection*{c. Generating properties of the constraints}
By using the algorithm, proposed in~\cite{Castellani}, we can construct a linear combination of the first class constraints, which generates our gauge transformations,
\be
\delta_f A_{\mu}(x) = \{G_f,A_{\mu}(x)\}_{\mathrm{p.s.}}.
\ee
 The answer reads,
\be
G_f := \int_{\Sigma(x^0)} \dd y\,\big(-\dot{f}(y)\phi^1(y) + f(y)\{A_0(y), \phi^1(y)\} + f(y)\, \phi^2(y)\big).
\ee
By a straightforward calculation one can check that these generators close the following algebra,
\be
 \{G_f,G_g\}_{\mathrm{p.s.}} =   G_{\{f,g\}}.
\ee

\section{Conclusions} \label{Conclu}
We analysed the Hamiltonian constraints for a generic Lie-Poisson gauge model with a purely spatial noncommutativity, which exhibits a Lagrangian formulation. 

On the one hand, we have seen that for each spatial point there are two first class constraints $\phi^1(x)$ and $\phi^2(x)$, defined by Eq.~\eqref{firsconstr} and Eq.~\eqref{seccostr} respectively, whilst the second class constraints are absent. From this point of view the situation is identical to the Maxwell theory, in particular,  the numbers of physical d.o.f. of the deformed and the undeformed models coincide.

On the other hand, the constraint algebra\footnote{The poisson bracket, appearing in the right-hand side of this equation, acts on the variable $x$.} 
\begin{eqnarray}
\{\phi^1(x),\phi^1(y)\}_{\mathrm{p.s.}} &=&0,  \nonumber\\
 \{\phi^1(x),\phi^2(y)\}_{\mathrm{p.s.}} &=&0, \nonumber\\
\{\phi^2(x),\phi^2(y)\}_{\mathrm{p.s.}} &=& -\{\phi^2(x),\delta^{n-1}(x-y)\},
\end{eqnarray}
is \emph{not} identical to its commutative counterpart, where all the constraints Poisson-commute strongly. A similar situation takes place for the Yang-Mills fields, see e.g.~\cite{Castellani}.

\label{sectconclusions}

 \vspace{5pt}

\begin{appendix}

{\section{Proof of the identity~\eqref{Notiden}}\label{appA}
Below we perform the standard calculation, which yields Noether identities. 
The invariance of the action~\eqref{dm1action} upon the infinitesimal transformations~\eqref{eqgaugetransfgeneralDM1} implies that for 
any $\tilde{f} \in\mathcal{C}^{\infty}\big(\Sigma(x^0)\big)$,
\begin{eqnarray}
0 {} & =&\delta_{\tilde{f}} S_{n-1}= \int_{\Sigma(x^0)}\dd^{n-1} x\, \frac{\delta S_{n-1}}{\delta A_{j}(x)} \,\delta_{\tilde{f}} A_{j}(x) \nonumber\\
& =&\int_{\Sigma(x^0)}\dd^{n-1} x \, \frac{\delta S_{n-1}}{\delta A_{j}(x)}
\,Q_{j}^k(x)\, \partial_{k} \tilde{f}(x).
\end{eqnarray}
Integrating by parts and assuming that all the fields decay sufficiently fast at the spatial infinity, we get the relation
\be
\int_{\Sigma(x^0)}\dd^{n-1} x \, \tilde{f}(x)\,\partial_{k}\left[\frac{\delta S_{n-1}}{\delta A_{j}(x)}
Q_j^k(x) \right]  = 0 , \qquad \forall \tilde{f}\in \mathcal{C}^{\infty}\big(\Sigma(x^0)\big),
\ee
therefore the integrand must be  identically zero, and we arrive at the identity~\eqref{Notiden}. Q.E.D.

\section{Proof of the relation~\eqref{twotwocomm}}\label{appB}

\noindent{\bf Remark \#1}. 
 The distributional equality~\eqref{twotwocomm}} is valid iff
\be \label{dadem}
\int_{\Sigma(x^0) \times\Sigma(x^0)} \dd x \dd y\, \varphi(x,y)\,\{\phi^2(x),\phi^2(y)\}_{\mathrm{p.s}} =  \int_{\Sigma(x^0) \times\Sigma(x^0)} \dd x \dd y\, \varphi(x,y)\, \{ \delta(x-y), \phi^2(x)\} , 
\ee
for any test function $\varphi(x,y)$. From now on we assume that $\varphi \in \mathcal{K}\big(\Sigma(x^0) \times\Sigma(x^0)\big)$, where $\mathcal{K}$ denotes the space of  infinitely differentiable functions with bounded support. It is known that the following set of functions
\be
\sum_{k=0}^n \varphi_{1,k}(x)\,\varphi_{2,k}(y), \qquad  \varphi_{1,k},  \varphi_{2,k}\in  \mathcal{K}\big(\Sigma(x^0) \big),\quad n\in \mathbb{N},
\ee
is dense in  $\mathcal{K}\big(\Sigma(x^0) \times\Sigma(x^0)\big)$. Since both lefthand and righthand sides of~\eqref{twotwocomm} are continuous linear functionals on $\mathcal{K}\big(\Sigma(x^0) \times\Sigma(x^0)\big)$,  it is sufficient to demonstrate~\eqref{dadem} for test functions  of the structure\footnote{ We borrowed this logic from Sec. 5.1 of Ref.~\cite{Gelfand}.}
\be \label{phistru}
\phi(x,y) = \varphi_1(x)\,\varphi_2(y),  \qquad \forall  \varphi_1, \varphi_2 \in  \mathcal{K}\big(\Sigma(x^0) \big).
\ee

\noindent{\bf Remark \#2}  For any  $\tilde{f}, \tilde{g} \in  \mathcal{K}\big(\Sigma(x^0) \big)$ the following identity holds
\be \label{useful}
\int_{\Sigma(x^0) } \dd y\, Q_q^r(y)\, \frac{\delta Q_j^k (x)}{\delta A_q (y)}\,\left[ \partial_{r[y]}\tilde{g}(y)\,\partial_{k[x]}\tilde{f}(x)  - (\tilde{f} \leftrightarrow\tilde{g})\right] = Q_j^k(x) \, \partial_{k[x]} \{\tilde{f}(x), \tilde{g}(x)\},
\ee
where we used the  shorthand notations $ \partial_{k[x]} ={\partial}/{\partial x^k} $ and $\partial_{p[y]} ={\partial}/{\partial y^p}$. Indeed, this formula is nothing but the $n-1$-dimensional commutative relation~\eqref{Poisdm1}, rewritten in terms of $Q$ and its variational derivatives.

\noindent{\bf Proof}. By substituting Eq.~\eqref{phistru} in the lefthand side of the required relation~\eqref{dadem}, and by applying~\eqref{useful} at $\tilde{f} = \varphi_2$ and $\tilde{g} =\varphi_1$, we obtain
\begin{eqnarray}
&&\int_{\Sigma(x^0) \times\Sigma(x^0)} \dd x \dd y\, \varphi(x,y)\, \{\phi^2(x),\phi^2(y)\}_{\mathrm{p.s}}   \nonumber \\
&&=\int_{\Sigma(x^0) \times\Sigma(x^0)} \dd x \dd y\, \Pi^i(x)\,
Q_j^p(y)\frac{\delta Q_i^k(x)}{\delta A_j(y)} \, \bigg[\partial_{k[x]}\varphi_1(x)\,\partial_{p[y]}\,\varphi_2(y) - \big(\varphi_1\leftrightarrow\varphi_2 \big) \bigg]\nonumber\\
&&=\int_{\Sigma(x^0)} \dd x \,\phi^2(x)\, \{\varphi_1(x),\varphi_2(x)\}.
\end{eqnarray}
By using the compatibility relation $\partial_q\Theta^{pq} =0$,  we can represent the Poisson bracket in the last line of the previous equation as follows,
\be
\{\varphi_1(x),\varphi_2(x)\} = - \int_{\Sigma(x^0)} \dd y\, \partial_{p[x]}\Big(\Theta^{pq}(x)\,\varphi_1(x)\varphi_2(y)\Big) 
\,\partial_{q[y]}\delta^{n-1}(y-x).
\ee
Therefore
\be
\int_{\Sigma(x^0)} \dd x \,\phi^2(x)\, \{\varphi_1(x),\varphi_2(x)\}  = \int_{\Sigma(x^0) \times\Sigma(x^0)} \dd x \dd y\, \varphi(x,y)\, \{ \delta(x-y), \phi^2(x)\},
\ee
what coincides with the righthand side of the desired equality~\eqref{dadem}. Q.E.D.

\end{appendix}

\section*{Acknowledgments}
The Authors are grateful to Patrizia Vitale for having suggested the problem, addressed in the present paper.
M.K. is grateful to Vlad Kupriyanov and to Patrizia Vitale for collaboration on the related topics.

\end{document}